# Enhanced thermal stability of dielectric and energy storage properties in 0.4BCZT-0.6BTSn lead-free ceramics elaborated by sol-gel method

S. Khardazi[1*], H. Zaitouni[1], A. Neqali[1], S. Lyubchyk[2], D. Mezzane[1,3], M. Amjoud[1], E. Choukri[1], S. Lyubchyk[2], Z. Kutnjak[4]

[1] IMED-Lab, Cadi-Ayyad University, Faculty of Sciences and Technology, Department of Applied Physics, Marrakech, 40000, Morocco

[2] DeepTechLab, Faculty of Engineering, Lusófona University, 376 Campo Grande, 1749-024 Lisbon, Portugal

[3] Laboratory of Physics of Condensed Matter (LPMC), University of Picardie Jules Verne, Scientific Pole, 33 Rue Saint-Leu, Amiens Cedex 1 80039, France

[4] Jožef Stefan Institute, Ljubljana, 1000, Slovenia

**Abstract**

Polycrystalline lead-free $Ba_{0.85}Ca_{0.15}Zr_{0.10}Ti_{0.90}O_3$ (BCZT), $BaTi_{0.89}Sn_{0.11}O_3$ (BTSn) and $0.4Ba_{0.85}Ca_{0.15}Zr_{0.10}Ti_{0.90}O_3$–$0.6BaTi_{0.89}Sn_{0.11}O_3$ (0.4BCZT–0.6BTSn) ferroelectric ceramics were prepared via sol-gel process and their structural, dielectric and energy storage properties were studied. Pure perovskite structure was confirmed by X-ray diffraction analysis. The evolution of energy storage performances with temperature was studied. A Significant recoverable energy-storage density of 137.86 mJ/cm$^3$ and high energy-storage efficiency of 86.19% under a moderate electric field of 30 kV/cm were achieved in the composite 0.4BCZT–0.6BTSn ceramic at 353 K. Moreover, excellent temperature stability (70 – 130 °C) of the energy storage efficiency (less than 3%) was achieved.

*Corresponding author. IMED-Lab, Cadi-Ayyad University, Faculty of Sciences and Technology, Department of Applied Physics, Marrakech, 40000, Morocco.
E-mail address: khardazzisaid@gmail.com, said.khardazi@ced.uca.ma (S.Khardazi)
Tel: +212 608219671

## 1. Introduction

In the development of the electric industry, dielectric capacitors with excellent energy storage capability and energy efficiency are required [1]. On the other hand, the development of well-known eco-friendly lead-free perovskite ferroelectrics with dielectric, electrocaloric, electromechanical and energy storage properties, etc., has garnered a significant mounting of research interest [2]. Furthermore, perovskite materials with excellent structural stability and tolerance are considered attractive for exploring new families of multifunctional materials by chemical modification [3]–[5].

Antiferroelectric (AFE) materials have a double P-E loop, a tiny remnant polarization ($P_r$) and relatively high dielectric breakdown strength (BDS) at the morphotropic phase boundary(MBP), making them particularly promising for high-recoverable energy storage devices [1], [6], [7]. Unfortunately, frequent studies on the energy storage performances of AFE materials have been carried out on lead-based materials [8]–[10]. However, due to the toxicity of lead compounds, several research projects have focused on finding new lead-free ferroelectric materials. Barium titanate $BaTiO_3$ (BT) is one of the most environment-friendly lead-free materials that has been explored for applications such as piezoelectric transducers, sensors, refrigeration, and capacitors [11], [12]. In addition, the substitution in site A or/ and B is one of the approaches to develop the best physical properties in BT and push the phase transition boundary close to the room temperature (RT). The substitution of $Ti^{4+}$ by $Sn^{4+}$ into the B-site of BT leads to $BaTi_{1-x}Sn_xO_3$ (BTSn) system whose phase is directly impacted by the Sn concentration. Moreover, as the Sn content rises, the ceramic's dielectric permittivity notably increases and exceeds that of pure BT as well as the $T_c$ shifts to room temperature (RT) [13], [14]. To our best knowledge, $BaTi_{0.89}Sn_{0.11}O_3$ composition shows a high dielectric constant, giant piezoelectric coefficient ($d_{33}$~ 1100 $pCN^{-1}$), large electrocaloric effect and enhanced energy storage performances due to existence of quasi-quadruple point near room temperature [15]–[20]. Likewise, it has been reported that $Ba_{0.85}Ca_{0.15}Zr_{0.10}Ti_{0.90}O_3$ (BCZT), one of the BT-based materials, features a high piezoelectric coefficient ($d_{33}$~620 $pCN^{-1}$), which is an even higher than that observed for the PZT system (with $d_{33}$ = 500–600 pC/N) due to the existence of MBP region [21], [22]. Hence, the development of BCZT relaxor ceramic has prompted a rise in interest in energy storage capacities [23]–[25].

Subsequently, the solid solution of BCZT and BTSn ($0.4Ba_{0.85}Ca_{0.15}Zr_{0.10}Ti_{0.90}O_3$–$0.6BaTi_{0.89}Sn_{0.11}O_3$) could be an effective way to improve the energy storage through the creation of multi-phases approach. This approach has been employed in several BT-based systems [26]–[30]. Based on the previous work on the $(1-x)Ba_{0.85}Ca_{0.15}Zr_{0.10}Ti_{0.90}O_3$–

$x$BaTi$_{0.89}$Sn$_{0.11}$O$_3$ ceramics synthesized by solid state method, the composition $x = 0.6$ exhibits improved dielectric and ferroelectric properties [31].

In this context, the present work explores the energy storage properties of the designed solid solution 0.4Ba$_{0.85}$Ca$_{0.15}$Zr$_{0.10}$Ti$_{0.90}$O$_3$–0.6BaTi$_{0.89}$Sn$_{0.11}$O$_3$ (named as 0.4BCZT–0.6BTSn). The samples were prepared by the sol-gel route, and their structural, dielectric, ferroelectric and energy storage performances were investigated.

## 2. Experimental Procedure

The polycrystalline ceramics Ba$_{0.85}$Ca$_{0.15}$Zr$_{0.10}$Ti$_{0.90}$O$_3$ (BCZT) and BaTi$_{0.89}$Sn$_{0.11}$O$_3$ (BTSn) were prepared individually by sol-gel technique and then mixed in a suitable weight ratio to acquire the desired composite ceramic 0.4Ba$_{0.85}$Ca$_{0.15}$Zr$_{0.10}$Ti$_{0.90}$O$_3$ –0.6 BaTi$_{0.89}$Sn$_{0.11}$O$_3$ (0.4BCZT–0.6BTSn). The chemical reagents barium acetate Ba(CH$_3$COO)$_2$, tin chloride dehydrate SnCl$_2$-2H$_2$O, titanium Isopropoxide (C$_{12}$H$_{28}$O$_4$Ti), zirconium oxychloride (ZrOCl$_2$·8H$_2$O) and calcium nitrate tetrahydrate (Ca(NO$_3$)$_2$·4H$_2$O) were used as initial precursors. Acetic acid CH$_3$CH$_2$OOH and 2- methoxyethanol (C$_3$H$_8$O$_2$) were used as solvent agents. The BCZT powder synthesis process details can be found in ref [32]. All reagents were of analytical grade and used as received. For the synthesis procedure of BTSn, a stoichiometric quantity of barium acetate (Ba(CH3- COO)$_2$) was dissolved in glacial acetic acid, while tin chloride (SnCl$_2$.2H$_2$O) was dissolved separately in 2-methoxyethanol. The two solutions were then mixed at room temperature. Finally, the C$_{12}$H$_{28}$O$_4$Ti was added to the mixture. The pH of the solution was adjusted to 6 by adding ammonia (NH$_4$OH). The obtained solution was heated at 80 °C for 1h leading to a transparent gel. Then, the gels BCZT and BTSn were washed with distilled water and ethanol and were separately calcined at 1000 °C for 5h. Subsequently, the pure phases of BCZT and BTSn powders were mixed with a mole ratio of 1:3 for 1h with absolute ethanol in an agate mortar and then dried overnight. Circular pellets with 6 mm diameter and 0.4 mm thickness were fabricated using a uniaxial hydraulic press and then sintered at 1350 °C for 7 h.

The room temperature (RT) crystalline structure of the sintered BCZT, BTSn and 0.4BCZT–0.6BTSn ceramics was investigated by X-ray diffraction (XRD) using Panalytical X-Pert Pro under Cu-Kα radiation with λ ~ 1.540598 Å. The samples were characterized for microstructure using a TESCAN VEGA3 Scanning Electron Microscope (SEM) and the grain size distributions were determined using ImageJ softwar. The dielectric measurements were performed in the frequency range 100 Hz – 1 MHz and temperature interval from 20 °C to 250 °C, using an impedance meter HP 4284A.

Temperature-dependent P–E hysteresis curves were measured at 200 Hz using a ferroelectric test system (PolyK Technologies State College, PA, USA).

## 3. Results and discussion

3.1 X-ray diffraction analysis

The X-ray diffraction patterns of BCZT, BTSn and 0.4BCZT–0.6BTSn sintered ceramics were recorded at RT in the range 20–90° as shown in Fig. 1. All ceramics possess a pure perovskite structure without any secondary phase. All the diffraction peaks are indexed with reference to the standard powder diffraction patterns of the $BaTiO_3$ perovskite structure. The most common diffraction peaks of barium titanate around $2\theta = 45°$ are the $(200)_R$ of the rhombohedral phase, $(022)/(200)_O$ of the orthorhombic phase and $(002)/(200)_T$ of the tetragonal phase [33]. Therefore, the Rietveld refinement is essential for analyzing the structural properties. Using the Fullprof software, the XRD patterns of BTSn, BCZT, and 0.4BCZT-0.6BTSn samples were refined using the Rietveld method(Figure. 2 a-c). The crystal structure of the BTSn composition was successfully refined in the tetragonal (T) phase of the *P4mm* space group. The BCZT sample was refined using a mixture of tetragonal (P4*mm*) and orthorhombic (O) (*Pmm*2) phases attributed to $(022)_O$, $(200)_T$, and $(200)_O$ reflection peaks. There are numerous reports with coexistence phases at room temperature for similar compositions characterizing the formation of the Morphotropic Phase Boundary (MPB) in the BCZT sample [34], [35]. For the 0.4BCZT-0.6BTSn sample, the phase analysis was carried out using a combination of (T) (P*4mm*) and (O) (*Amm*2) symmetries that coexisted at room temperature. Indeed, the coexistence of (T) and (O) symmetries in the designed 0.4BCZT-0.6BTSn is consistent with previous reports [31]. Rietveld refined fit parameters for all compositions are gathered in Table 1.

To get insight into the microstrain generated due to the combination of BCZT and BTSn phases, Williamson-Hall (W–H) method was adopted using the following equation [36]:

$$\beta_T cos\theta = \frac{K\lambda}{D} + 4\varepsilon \, sin\theta. \qquad (1)$$

Where the parameters K, D, θ, λ and ε denote the particle shape factor (~0.94), effective crystallite size, diffraction angle, the wavelength of the X-ray source and the effective strain generated in the samples, respectively. Fig. 2d depicts the plots of $\beta_T cos\theta$ vs. $sin\theta$ for BTSn, BCZT, and 0.4BCZT-0.6BTSn samples. The microstrain parameter of the different samples is presented in Table 1. One can see that the strain in the 0.4BCZT-0.6BTSn is found to be intermediate between that of the pure BTSn and BCZT samples. Based on the W-H method, the strain values obtained are $1.2\times10^{-3}$, $2.1\times10^{-3}$, and $1.8\times10^{-3}$ for BTSn, BCZT, and 0.4BCZT-0.6BTSn, respectively.

**(Insert Figure 1, here)**

**(Insert Figure 2, here)**

.

**(Insert Table 1, here)**

3.2 Microstructure analysis

The scanning electron microscopy (SEM) micrographs and grain size distribution of all the investigated compositions are illustrated in Fig. 3. The pure BCZT and BTSn ceramics shown in Fig. 2a and 2b, respectively, exhibit non-uniform grain size. Besides, the designed composition 0.4BCZT–0.6BTSn displays an homogenous grain distribution and dense microstructure, which could enhance its breakdown strength (BDS). The average grain size of the BCZT composition was found to be around $17.57\pm 0.17$ µm which is larger than that of the BTSn composition ($13.24\pm 0.15$ µm). For the composite solid solution 0.4BCZT–0.6BTSn, an average grain size of $11.72\pm 0.12$ µm is obtained, smaller than the simple compositions. The decrease in grain size in the composite solid solution could be ascribed to grain growth inhibition of the BCZT phase, where the BTSn phase acts as a grain growth inhibitor [37], [38]. The density of ceramics was determined using Archimede's method and was found to be 5.45, 5.94, and 5.87 g/cm$^3$ for BCZT, BTSn, and 0.4BCZT–0.6BTSn ceramics, respectively.

**(Insert Figure 3, here)**

3.3 Dielectric study

**(Insert Figure 4, here)**

In order to investigate the phase transition in the three studied samples, the temperature dependence of the dielectric constant and the dielectric loss (tgδ) for all the investigated compositions at various frequencies are depicted in Fig. 4. All the studied samples present a dielectric anomaly at high temperatures (around Curie temperature ($T_C$)) characterizing the Tetragonal–Cubic (T–C) phase transition, which denotes the Ferroelectric–Paraelectric (FE–PE) structural change. Moreover, it can be seen that for both samples BCZT and 0.4BCZT–0.6BTSn, the dielectric curves reveal the existence of another anomaly below the (T–C) phase transition corresponding to the Orthorhombic–Tetragonal (O-T) phase transition around RT. These results corroborate those obtained by XRD, supporting the coexistence of orthorhombic and tetragonal phases in both ceramics at room temperature. It is worth noting that the solid solution 0.4BCZT–0.6BTSn has a high maximum dielectric permittivity ($\varepsilon_m$) of 22190 at 1kHz near RT. The Obtained value is significantly greater than the value reported in the literature for the same composition prepared using the solid-state method [31]. The $\varepsilon_m$ value found for the composite compound is intermediate between that of BTSn and BCZT.

On the other hand, the temperature dependence of the dielectric loss tgδ exhibits a similar merging tendency of dielectric anomalies, as shown in Fig. 3. All investigated compositions exhibit low dielectric losses, particularly the composite sample 0.4BCZT–0.6BTSn, which has small losses (tgδ < 0.05, at 1kHz) around the transition temperature, indicating excellent dielectric properties.

To get additional details regarding the phase transition, the temperature dependence of the inverse dielectric constant was plotted at 1 kHz of frequency and fitted using the Curie – Weiss law (Fig. 5):

$$\frac{1}{\varepsilon} = \frac{(T-T_0)}{C}, \qquad (2)$$

where $\varepsilon$ is the real part of the dielectric constant, $T_0$ and C are the Curie-Weiss temperature, and Curie-Weiss constant, respectively.

As shown in Fig. 5, the experimental data are well-fitted using Eq. (2). All ceramics have Curie constant values on the order of $10^5$ K, which is consistent with the Curie constant of well-known displacive-type ferroelectric such as $BaTiO_3$ ($1.7\times10^5$ K). This suggests that a displacive phase transition drives the high-temperature paraelectric phase in all samples [39].

Uchino and Nomura suggested an empirical formula to describe the fluctuation of the dielectric constant with the temperature above $T_C$ for relaxors in order to further evaluate the degree of diffusion of the dielectric curves. This formula given by Eq.3 is an updated version of the Curie-Weiss equation [40]:

$$\frac{1}{\varepsilon} - \frac{1}{\varepsilon_m} = \frac{(T-T_m)^\gamma}{C}, \qquad (3)$$

where $\varepsilon_m$ represents the maximum dielectric constant at the transition temperature $T_m$, C and γ (degree of the transition diffuseness) are constants. For γ = 1, Eq. 3 fits a normal ferroelectric, while for γ = 2, Eq. 3 indicates an ideal relaxor and represents the so-called complete diffuse phase transition (DPT). However, intermediate values of γ between 1 and 2, characterize an ''incomplete'' DPT [41]. Figures. 4(b), (d) and (f) show the plot of ln $(1/\varepsilon - 1/\varepsilon_m)$ versus ln(T - $T_m$) at 1 kHz. Fitting the experimental data with Eq.(2) results in γ values of 1.73, 1.57 and 1.54 for BCZT, BTSn, and 0.4BCZT–0.6BTSn samples, respectively. According to this result, all samples exhibit classic ferroelectric behavior with an incomplete diffuse phase character.

**(Insert Figure 5, here)**

3.4 Ferroelectric and energy storage properties

The ferroelectric properties of all investigated compositions (BCZT, BTSn and 0.4BCZT–0.6BTSn, were examined by measuring the thermal evolution of P–E hysteresis loops near the Curie temperature using a maximum electric field of 30 kV/cm at 200 Hz. The curves are depicted in Figure 6(a-c). Each of the examined samples has been found to have typical ferroelectric P-E hysteresis loops. Indeed, when temperature rises, the hysteresis curve becomes thinner and changes to a linear response as paraelectric domains appear, indicating that each composition has undergone a FE-PE phase transition. Fig. 6(d) illustrates the P–E hysteresis loops of the BCZT, BTSn and 0.4BCZT–0.6BTSn samples at RT. In contrast to the pure BTSn and BCZT samples, the P-E loop of the designed solid solution 0.4BCZT-0.6BTSn displays a highly saturated polarization and a slimmer form. The remnant polarization Pr for pure BTSn and BCZT was 10.09 µC/cm$^2$ and 8.45 µC/cm$^2$, respectively, whereas the 0.4 BCZT-0.6 BTSn ceramic exhibits the lowest value at 7.54 µC/cm$^2$. As shown in Fig. 5(e), the average coercive field ($E_C$) values slightly decreased from 3.60 kV/cm for pure BCZT to 1.35 kV/cm for 0.4BCZT–0.6BTSn. The gradual reduction in $P_r$ in the 0.4BCZT–0.6BTSn sample could be attributable to smaller grains, which can be explained by an increase in the clamping effect from grain boundaries, as previously seen in SEM micrographs [42]. Moreover, the $E_C$ in the 0.4BCZT-0.6BTSn sample significantly decreased, which suggests that the ferroelectric domains can switch polarizations more easily [43]. These findings suggest that the designed composite 0.4BCZT–0.6BTSn may offer enhanced energy storage capabilities.

As well known, ferroelectric materials with high dielectric constants and tiny hysteresis loops may exhibit improved energy storage capabilities. To get an insight into these performances, all of the studied samples 'P-E curves' temperature dependency was recorded under an electric field of 30 kV/cm, as shown in Fig. 6. (a-c). Typically, the electrical energy storage capabilities (recoverable energy density, total energy density, etc.) of dielectric materials can be derived from their P–E loops. For example the, recoverable energy density ($W_{rec}$) is obtained by integrating the area between the polarization axis and the upper branch of the P–E curve (blue area in Fig. 7. The equations (4) and (5) were used to calculate each of $W_{rec}$ and energy storage efficiency ($\eta$) [44]:

$$W_{rec} = \int_{Pr}^{Pmax} EdP \qquad (4)$$

$$\eta(\%) = \frac{W_{rec}}{W_{tot}} * 100 = \frac{W_{rec}}{W_{rec}+W_{loss}} * 100 \qquad (5)$$

$P_{max}$ is the maximal polarization, and $W_{tot}$ is the total energy density. Fig. 8(a) depicts the thermal evolution of recoverable energy density $W_{rec}$ of the three investigated samples. One can see that $W_{rec}$ increases with temperature and reaches a maximum around the Curie temperature, which corresponds to the maximum of permittivity. The highest computed $W_{rec}$ are around 99.88 mJ/cm$^3$ at 100.71 °C, 99.83 mJ/cm$^3$ at 44.75 °C and 137.86 mJ/cm$^3$ at 80°C for BCZT, BTSn and 0.4BCZT–0.6BTSn ceramics, respectively at the maximum electric field. Analogous to $W_{rec}$, the energy loss density $W_{loss}$ (depicted by the red area in Fig. 7) are about 44.95, 79.75 and 26.77 mJ/cm$^3$ for BCZT, BTSn and 0.4BCZT–0.6BTSn ceramics, respectively. The temperature dependence of the energy storage efficiency ($\eta$) for all elaborated ceramics is shown in Fig. 8(b). Under the same applied electric field of 30 kV/cm, the 0.4BCZT–0.6BTSn solid solution ceramic exhibits a higher energy storage efficiency (86.19%) compared to the BCZT (74.12%) and BTSn (55.73%) pure ceramics. The coexistence of multiple phases not only increases the dielectric constant but also improves the energy storage density as demonstrated in previous works [18], [36], [45], [46]. As a result, the designed solid solution 0.4BCZT–0.6BTSn showed a high energy storage density and improved energy storage efficiency.

It should be noted that higher efficiency means less energy loss during the charge and discharge process. The above formula (Eq. 4) demonstrates that the more significant the difference between $P_{max}$ and $P_r$, the better the effective energy storage performances. Furthermore, the practical applications of electrical devices largely depend on the thermal stability of energy storage performance [47], [48]. Fig. 8-c depicts the thermal evolution of the energy storage efficiency of the 0.4BCZT–0.6BTSn sample. One can see that 0.4BCZT–0.6BTSn ceramic exhibits excellent stability of the energy storage efficiency of less than 3% ($\eta \approx 81.37 - 79.50$ %) in the temperature range of 70 – 130 °C.

To set out our results to the literature, Table 2 summarizes the comparison of the recoverable energy density and the energy storage efficiency of various lead-free ceramics. Notably, increasing the applied electric field induces higher polarization, which improving energy storage performances. Under the same applied electric field, our samples, particularly the 0.4BCZT–0.6BTSn ceramic, show a higher $W_{rec}$ and energy efficiency $\eta$ than previously reported works on $BaTiO_3$-based ceramics. For instance, Merselmiz *et al.* [21] found a $W_{rec}$ of 62 mJ/cm$^3$ and a $\eta$ of 72.9 % at 130 °C in $Ba_{0.85}Ca_{0.15}Zr_{0.10}Ti_{0.90}O_3$ ceramics synthesized by the solid-state method. For the same composition, $Ba_{0.85}Ca_{0.15}Zr_{0.10}Ti_{0.90}O_3$ ceramics prepared by SDS-assisted solvothermal synthesis, Hanani *et al.* [49] reported a $W_{rec}$ of 14 mJ/cm$^3$ and η of 80 % at 129 °C under a low applied electric field of 6.5 kV/cm. Moreover, under a high electric field of 90 kV/cm, Benyoussef *et al.* [50] found a high $W_{rec}$ of 1200 mJ/cm$^3$ and a $\eta$ of 64 % at an elevated temperature of 200 °C in $Na_{0.5}(Bi_{0.98}Dy_{0.02})_{0.5}TiO_3$ prepared via the solid-state method. Note that the performance of energy storage may be influenced by the elaboration methods, chemical composition, grain size engineering, and applied electric fields. According to the study's results, combining various phases together is a promising way to boost energy storage capacity.

**(Insert Figure 6, here)**

**(Insert Figure 7, here)**

**(Insert Figure 8, here)**

**(Insert Table 2, here)**

**Conclusion**

In the present paper, structural, dielectric, ferroelectric and energy storage properties of pure perovskite lead-free BCZT, BTSn and 0.6BTSn-0.4BCZT ferroelectric ceramics have been investigated. Rietveld refinement of XRD data confirms the coexistence of the rhombohedral and orthorhombic phases at room temperature in the composite 0.4BCZT–0.6BTSn ceramic. Remarkably, an improved recoverable energy density of 137.86 mJ/cm$^3$ and a high energy storage efficiency of 86.19 % at 80 °C under a moderate applied electric field of 30 kV/cm were achieved in the designed 0.4BCZT–0.6BTSn ceramic. Besides, the sample exhibits excellent thermal stability of the energy storage efficiency (less than 3%) in the temperature range of 70 to 130 °C under 30 kV/cm. Such results make the pb-free 0.4BCZT–0.6BTSn ferroelectric ceramic a very promising potential matrix for energy storage capacitor applications.

**Acknowledgements**

This research is financially supported by the European Union Horizon 2020 Research and Innovation actions MSCA-RISE-ENGIMA (No. 778072) and MSCA-RISE-MELON (No. 872631)

**References**

[1] Z. Liu, X. Dong, Y. Liu, F. Cao, et G. Wang, « Electric field tunable thermal stability of energy storage properties of PLZST antiferroelectric ceramics », *J Am Ceram Soc*, vol. 100, n° 6, p. 2382-2386, juin 2017, doi: 10.1111/jace.14867.

[2] D. Zheng, R. Zuo, D. Zhang, et Y. Li, « Novel BiFeO$_3$-BaTiO$_3$-Ba(Mg$_{1/3}$Nb$_{2/3}$)O$_3$ Lead-Free Relaxor Ferroelectric Ceramics for Energy-Storage Capacitors », *J. Am. Ceram. Soc.*, vol. 98, n° 9, p. 2692-2695, sept. 2015, doi: 10.1111/jace.13737.

[3] K. P. Goetz, A. D. Taylor, F. Paulus, et Y. Vaynzof, « Shining Light on the Photoluminescence Properties of Metal Halide Perovskites », *Adv. Funct. Mater.*, vol. 30, n° 23, p. 1910004, juin 2020, doi: 10.1002/adfm.201910004.

[4] Z. Sun, Z. Wang, Y. Tian, G. Wang, W. Wang, M. Yang, X. Wang, F. Zhang, et Y. Pu, « Progress, Outlook, and Challenges in Lead-Free Energy-Storage Ferroelectrics », *Adv. Electron. Mater.*, vol. 6, nº 1, p. 1900698, janv. 2020, doi: 10.1002/aelm.201900698.

[5] Y. Dong, Y. Zhang, X. Li, Y. Feng, H. Zhang, et J. Xu, « Chiral Perovskites: Promising Materials toward Next-Generation Optoelectronics », *Small*, vol. 15, nº 39, p. 1902237, sept. 2019, doi: 10.1002/smll.201902237.

[6] Z. Liu, J. Lu, Y. Mao, P. Ren, et H. Fan, « Energy storage properties of $NaNbO_3$-$CaZrO_3$ ceramics with coexistence of ferroelectric and antiferroelectric phases », *Journal of the European Ceramic Society*, vol. 38, nº 15, p. 4939-4945, déc. 2018, doi: 10.1016/j.jeurceramsoc.2018.07.029.

[7] Z. Liu, X. Chen, W. Peng, C. Xu, X. Dong, F. Cao, et G. Wang, « Temperature-dependent stability of energy storage properties of $Pb_{0.97}La_{0.02}(Zr_{0.58}Sn_{0.335}Ti_{0.085})O_3$ antiferroelectric ceramics for pulse power capacitors », *Appl. Phys. Lett.*, vol. 106, nº 26, p. 262901, juin 2015, doi: 10.1063/1.4923373.

[8] F. Zhuo, Q. Li, Y. Li, J. Gao, Q. Yan, Y. Zhang, X. Xi, X. Chu, et W. Cao, « Field induced phase transitions and energy harvesting performance of $(Pb,La)(Zr,Sn,Ti)O_3$ single crystal », *Journal of Applied Physics*, vol. 121, nº 6, p. 064104, févr. 2017, doi: 10.1063/1.4975786.

[9] J. Wei, T. Yang, et H. Wang, « Excellent Energy Storage and Charge-discharge Performances in $PbHfO_3$ Antiferroelectric Ceramics », *Journal of the European Ceramic Society*, vol. 39, nº 2-3, p. 624-630, févr. 2019, doi: 10.1016/j.jeurceramsoc.2018.09.039.

[10] F. Bian, S. Yan, C. Xu, Z. Liu, X. Chen, C. Mao, F. Cao, J. Bian, G. Wang, et X. Dong, « Enhanced breakdown strength and energy density of antiferroelectric $Pb,La(Zr,Sn,Ti)O_3$ ceramic by forming core-shell structure », *Journal of the European Ceramic Society*, vol. 38, nº 9, p. 3170-3176, août 2018, doi: 10.1016/j.jeurceramsoc.2018.03.028.

[11] H. Abdmouleh, I. Kriaa, N. Abdelmoula, Z. Sassi, et H. Khemakhem, « The effect of $Zn^{2+}$ and $Nb^{5+}$ substitution on structural, dielectric, electrocaloric properties, and energy storage density of $Ba_{0.95}Ca_{0.05}Ti_{0.95}Zr_{0.05}O_3$ ceramics », *Journal of Alloys and Compounds*, vol. 878, p. 160355, oct. 2021, doi: 10.1016/j.jallcom.2021.160355.

[12] G. Dai, S. Wang, G. Huang, G. Chen, B. Lu, D. Li, T. Tao, Y. Yao, B. Liang, et S.G. Lu, « Direct and indirect measurement of large electrocaloric effect in barium strontium titanate ceramics », *Int J Appl Ceram Technol*, vol. 17, nº 3, p. 1354-1361, mai 2020, doi: 10.1111/ijac.13384.

[13]   Z. Luo, D. Zhang, Y. Liu, D. Zhou, Y. Yao, C. Liu, B. Dkhil, X. Ren, et X. Lou, « Enhanced electrocaloric effect in lead-free BaTi$_{1-x}$Sn$_x$O$_3$ ceramics near room temperature », *Appl. Phys. Lett.*, vol. 105, nº 10, p. 102904, sept. 2014, doi: 10.1063/1.4895615.

[14]   S. K. Upadhyay, V. R. Reddy, P. Bag, R. Rawat, S. M. Gupta, et A. Gupta, « Electrocaloric effect in lead-free Sn doped BaTiO$_3$ ceramics at room temperature and low applied fields », *Appl. Phys. Lett.*, vol. 105, nº 11, p. 112907, sept. 2014, doi: 10.1063/1.4896044.

[15]   Y. Yao, C. Zhou, D. Lv, D. Wang, H. Wu, Y. Yang, et X. Ren, « Large piezoelectricity and dielectric permittivity in BaTiO$_3$-xBaSnO$_3$ system: The role of phase coexisting », *EPL*, vol. 98, nº 2, p. 27008, avr. 2012, doi: 10.1209/0295-5075/98/27008.

[16]   M. Sanlialp, Z. Luo, V. Shvartsman, X. Wei, Y. Liu, B. Dkhil, et D.C. Lupascu, « Direct measurement of electrocaloric effect in lead-free Ba(Sn$_x$Ti$_{1-x}$)O$_3$ ceramics », *Appl. Phys. Lett.*, vol. 111, nº 17, p. 173903, oct. 2017, doi: 10.1063/1.5001196.

[17]   S. Merselmiz, Z. Hanani, D. Mezzane, M. Spreitzer, A. Bradeško, D. Fabijan, D. Vengust, M. Amjoud, L. Hajji, Z. Abkhar, A.G. Razumnaya, B. Rožič, et I.A. Luk'yanchuk, « High energy storage efficiency and large electrocaloric effect in lead-free BaTi$_{0.89}$Sn$_{0.11}$O$_3$ ceramic », *Ceramics International*, vol. 46, nº 15, p. 23867-23876, oct. 2020, doi: 10.1016/j.ceramint.2020.06.163.

[18]   J. Gao, Y. Wang, Y. Liu, X. Hu, X. Ke, L. Zhong, Y. He, et X. Ren, « Enhancing dielectric permittivity for energy-storage devices through tricritical phenomenon », *Sci Rep*, vol. 7, nº 1, p. 40916, mars 2017, doi: 10.1038/srep40916.

[19]   C. Zhao, J. Yang, Y. Huang, X. Hao, et J. Wu, « Broad-temperature-span and large electrocaloric effect in lead-free ceramics utilizing successive and metastable phase transitions », *J. Mater. Chem. A*, vol. 7, nº 44, p. 25526-25536, 2019, doi: 10.1039/C9TA10164K.

[20]   M. Zahid, Y. Hadouch, M. Amjoud, D. Mezzane, M. Gouné, K. Hoummada, A. Alimoussa, A.G. Razumnaya, B. Rožič, et Z. Kutnjak, « Enhanced near-ambient temperature energy storage and electrocaloric effect in the lead-free BaTi$_{0.89}$Sn$_{0.11}$O$_3$ ceramic synthesized by sol–gel method », *J Mater Sci: Mater Electron*, vol. 33, nº 16, p. 12900-12911, juin 2022, doi: 10.1007/s10854-022-08233-6.

[21]   S. Merselmiz, Z. Hanani, D. Mezzane, A.G. Razumnaya, M. Amjoud, L. Hajji, S. Terenchuk, B. Rožič, I.A. Luk'yanchuk, et Z. Kutnjak, « Thermal-stability of the enhanced piezoelectric, energy storage and electrocaloric properties of a lead-free BCZT ceramic », *RSC Adv.*, vol. 11, nº 16, p. 9459-9468, 2021, doi: 10.1039/D0RA09707A.

[22]   W. Liu et X. Ren, « Large Piezoelectric Effect in Pb-Free Ceramics », *Phys. Rev. Lett.*, vol. 103, nº 25, p. 257602, déc. 2009, doi: 10.1103/PhysRevLett.103.257602.

[23]   P. Yang, L. Li, S. Yu, W. Peng, et K. Xu, « Ultrahigh and field-independent energy storage efficiency of $(1-x)(Ba_{0.85}Ca_{0.15})(Zr_{0.1}Ti_{0.9})O_3-xBi(Mg_{0.5}Ti_{0.5})O_3$ ceramics », *Ceramics International*, vol. 47, nº 3, p. 3580-3585, févr. 2021, doi: 10.1016/j.ceramint.2020.09.206.

[24]   X. Chen, X. Chao, et Z. Yang, « Submicron barium calcium zirconium titanate ceramic for energy storage synthesised via the co-precipitation method », *Materials Research Bulletin*, vol. 111, p. 259-266, mars 2019, doi: 10.1016/j.materresbull.2018.11.025.

[25]   H. Mezzourh, S. Belkhadir, D. Mezzane, M. Amjoud, E. Choukri, A. Lahmar, Y. Gagou, Z. Kutnjak, et M. El Marssi, « Enhancing the dielectric, electrocaloric and energy storage properties of lead-free $Ba_{0.85}Ca_{0.15}Zr_{0.1}Ti_{0.9}O_3$ ceramics prepared via sol-gel process », *Physica B: Condensed Matter*, vol. 603, p. 412760, févr. 2021, doi: 10.1016/j.physb.2020.412760.

[26]   X. Chen, X. Li, J. Sun, C. Sun, J. Shi, F. Pang, et H. Zhou, « Simultaneously achieving ultrahigh energy storage density and energy efficiency in barium titanate based ceramics », *Ceramics International*, vol. 46, nº 3, p. 2764-2771, févr. 2020, doi: 10.1016/j.ceramint.2019.09.265.

[27]   J. Lv, Q. Li, Y. Li, M. Tang, D. Jin, Y. Yan, B. Fan, L. Jin, et G. Liu, « Significantly improved energy storage performance of NBT-BT based ceramics through domain control and preparation optimization », *Chemical Engineering Journal*, vol. 420, p. 129900, sept. 2021, doi: 10.1016/j.cej.2021.129900.

[28]   S. Song, Y. Jiao, F. Chen, X. Zeng, X. Wang, S. Zhou, T. Ai, G. Liu, et Y. Yan, « Ultrahigh electric breakdown strength, excellent dielectric energy storage density, and improved electrocaloric effect in Pb-free $(1-x)Ba(Zr_{0.15}Ti_{0.85})O_3-xNaNbO_3$ ceramics », *Ceramics International*, vol. 48, nº 8, p. 10789-10802, avr. 2022, doi: 10.1016/j.ceramint.2021.12.295.

[29]   C. Shi, F. Yan, G. Ge, Y. Wei, J. Zhai, et W. Yao, « Significantly enhanced energy storage performances and power density in $(1 - x)$BCZT-xSBT lead-free ceramics via synergistic optimization strategy », *Chemical Engineering Journal*, vol. 426, p. 130800, déc. 2021, doi: 10.1016/j.cej.2021.130800.

[30]   S. Luo, D.-Y. Zheng, C. Zhang, Y. Zhang, et B. Li, « Effects of BLT doping on electrical properties and relaxation behavior of BCZT-BLT ceramics », *J Mater Sci: Mater Electron*, vol. 31, nº 23, p. 21005-21016, déc. 2020, doi: 10.1007/s10854-020-04614-x.

[31]   S. Merselmiz, Z. Hanani, U. Prah, D. Mezzane, L. Hajji, Z. Abkhar, M. Spreitzer, D. Vengust, H. Uršič, D. Fabijan, A.G. Razumnaya, O. Shapovalova, I.A. Luk'yanchuk, et Z. Kutnjak, « Design of lead-free BCZT-based ceramics with enhanced piezoelectric energy harvesting performances », *Phys. Chem. Chem. Phys.*, vol. 24, nº 10, p. 6026-6036, 2022, doi: 10.1039/D1CP04723J.

[32]   S. Belkhadir, S. Ben Moumen, B. Asbani, M. Amjoud, D. Mezzane, I.A. Luk'yanchuk, E. Choukri, L. Hajji, Y. Gagou, et M. El Marssi, « Impedance spectroscopy analysis of the diffuse phase transition in lead-free $(Ba_{0.85}Ca_{0.15})(Zr_{0.1}Ti_{0.9})O_3$ ceramic elaborated by sol-gel method », *Superlattices and Microstructures*, vol. 127, p. 71-79, mars 2019, doi: 10.1016/j.spmi.2018.03.009.

[33]   I. Coondoo, N. Panwar, D. Alikin, I. Bdikin, S.S. Islam, A. Turygin, V.Y. Shur, et A.L. Kholkin, « A comparative study of structural and electrical properties in lead-free BCZT ceramics: Influence of the synthesis method », *Acta Materialia*, vol. 155, p. 331-342, août 2018, doi: 10.1016/j.actamat.2018.05.029.

[34]   W. Cai, Q. Zhang, C. Zhou, R. Gao, F. Wang, G. Chen, X. Deng, Z. Wang, N. Deng, L. Cheng, et C. Fu, « Effects of oxygen partial pressure on the electrical properties and phase transitions in $(Ba,Ca)(Ti,Zr)O_3$ ceramics », *J Mater Sci*, vol. 55, nº 23, p. 9972-9992, août 2020, doi: 10.1007/s10853-020-04771-8.

[35]   Z. Hanani, D. Mezzane, M. Amjoud, S. Fourcade, A.G. Razumnaya, I.A. Luk'yanchuk, et M. Gouné, « Enhancement of dielectric properties of lead-free BCZT ferroelectric ceramics by grain size engineering », *Superlattices and Microstructures*, vol. 127, p. 109-117, mars 2019, doi: 10.1016/j.spmi.2018.03.004.

[36]   R. L. Nayak, S. S. Dash, Y. Zhang, et M. P. K. Sahoo, « Enhanced dielectric, thermal stability, and energy storage properties in compositionally engineered lead-free ceramics at morphotropic phase boundary », *Ceramics International*, vol. 47, nº 12, p. 17220-17233, juin 2021, doi: 10.1016/j.ceramint.2021.03.033

[37]   T. Wang, J. Ma, B. Wu, F. Wang, S. Wang, M. Chen, et W. Wu, « Structure and Electrical Properties of Microwave Sintered BTS-BCT-xBF Lead-Free Piezoelectric Ceramics », *Materials*, vol. 15, nº 5, p. 1789, févr. 2022, doi: 10.3390/ma15051789.

[38]   K. Zhang, P. Gao, C. Liu, X. Chen, X. Huang, Y. Pu, et Z. Liu, « Structural Evolution and Enhanced Piezoelectric Activity in Novel Lead-Free $BaTiO_3$-$Ca(Sn_{1/2}Zr_{1/2})O_3$ Solid Solutions », *Energies*, vol. 15, nº 20, p. 7795, oct. 2022, doi: 10.3390/en15207795

[39]   P. Bharathi et K. B. R. Varma, « Grain and the concomitant ferroelectric domain size dependent physical properties of $Ba_{0.85}Ca_{0.15}Zr_{0.1}Ti_{0.9}O_3$ ceramics fabricated using

powders derived from oxalate precursor route », *Journal of Applied Physics*, vol. 116, n° 16, p. 164107, oct. 2014, doi: 10.1063/1.4900494.

[40] K. Uchino et S. Nomura, « Critical exponents of the dielectric constants in diffused-phase-transition crystals », *Ferroelectrics*, vol. 44, n° 1, p. 55-61, avr. 1982, doi: 10.1080/00150198208260644.

[41] A. Belboukhari, Z. Abkhar, E. Choukri, Y. Gagou, N. Abdelmoula, R. Elmoznine, D. Mezzane, H. Khemakhem, M. El Marssi, A.G Razumnaya, I. Raevski, et I.A. Luk'yanchuk, « Studies of Diffuse Phase Transition in Ferroelectric Solid Solution $Pb_{1-x}K_{2x}Nb_2O_6$ (x = 0.1, 0.2, 0.25 and 0.3) », *Ferroelectrics*, vol. 444, n° 1, p. 116-124, janv. 2013, doi: 10.1080/00150193.2013.786619.

[42] F. Q. Guo, B. H. Zhang, Z. X. Fan, X. Peng, Q. Yang, Y. X. Dong, et R. R. Chen, « Grain size effects on piezoelectric properties of $BaTiO_3$ ceramics prepared by spark plasma sintering », *J Mater Sci: Mater Electron*, vol. 27, n° 6, p. 5967-5971, juin 2016, doi: 10.1007/s10854-016-4518-1.

[43] M. Benyoussef, T. Mura, S. Saitzek, F. Azrour, J.F. Blach, A. Lahmar, Y. Gagou, M. El Marssi, A. Sayede, et M. Jouiad, « Nanostructured $BaTi_{1-x}Sn_xO_3$ ferroelectric materials for electrocaloric applications and energy performance », *Current Applied Physics*, vol. 38, p. 59-66, juin 2022, doi: 10.1016/j.cap.2022.03.012.

[44] Q. Li, W. Zhang, C. Wang, L. Ning, C. Wang, Y. Wen, B. Hu, et H. Fan, « Enhanced energy-storage performance of $(1-x)(0.72Bi_{0.5}Na_{0.5}TiO_3-0.28Bi_{0.2}Sr_{0.7\square 0.1}TiO_3)-xLa$ ceramics », *Journal of Alloys and Compounds*, vol. 775, p. 116-123, févr. 2019, doi: 10.1016/j.jallcom.2018.10.092.

[45] Y. Chen, Y. Wang, D. Zhao, H. Wang, X. He, Q. Zheng, et D. Lin, « Enhanced energy storage properties and dielectric stabilities in BNT-based ceramics via multiphase and dielectric peak broadening engineering », *Materials Chemistry and Physics*, vol. 290, p. 126542, oct. 2022, doi: 10.1016/j.matchemphys.2022.126542.

[46] Y. Li, J. Chen, P. Cai, et Z. Wen, « An electrochemically neutralized energy-assisted low-cost acid-alkaline electrolyzer for energy-saving electrolysis hydrogen generation », *J. Mater. Chem. A*, vol. 6, n° 12, p. 4948-4954, 2018, doi: 10.1039/C7TA10374C.

[47] K. Xu, P. Yang, W. Peng, et L. Li, « Temperature-stable MgO-doped BCZT lead-free ceramics with ultra-high energy storage efficiency », *Journal of Alloys and Compounds*, vol. 829, p. 154516, juill. 2020, doi: 10.1016/j.jallcom.2020.154516.

[48] Y. Zhou, J. Chen, B. Yang, et S. Zhao, « Ultrahigh energy storage performances derived from the relaxation behaviors and inhibition of the grain growth in La doped $Bi_5Ti_3FeO_{15}$

films », *Chemical Engineering Journal*, vol. 424, p. 130435, nov. 2021, doi: 10.1016/j.cej.2021.130435.

[49] Z. Hanani, D. Mezzane, M. Amjoud, A.G Razumnaya, S. Fourcade, Y. Gagou, K. Hoummada, M. El Marssi, et M. Gouné, , « Phase transitions, energy storage performances and electrocaloric effect of the lead-free $Ba_{0.85}Ca_{0.15}Zr_{0.10}Ti_{0.90}O_3$ ceramic relaxor », *J Mater Sci: Mater Electron*, vol. 30, nº 7, p. 6430-6438, avr. 2019, doi: 10.1007/s10854-019-00946-5.

[50] M. Benyoussef, M. Zannen, J. Belhadi, B. Manoun, J.L. Dellis, M. El Marssi, et A. Lahmar, « Dielectric, ferroelectric, and energy storage properties in dysprosium doped sodium bismuth titanate ceramics », *Ceramics International*, vol. 44, nº 16, p. 19451-19460, nov. 2018, doi: 10.1016/j.ceramint.2018.07.182.

[51] D. K. Kushvaha, S. K. Rout, et B. Tiwari, « Structural, piezoelectric and highdensity energy storage properties of lead-free BNKT-BCZT solid solution », *Journal of Alloys and Compounds*, vol. 782, p. 270-276, avr. 2019, doi: 10.1016/j.jallcom.2018.12.196.

[52] K. S. Srikanth et R. Vaish, « Enhanced electrocaloric, pyroelectric and energy storage performance of $BaCe_xTi_{1-x}O_3$ ceramics », *Journal of the European Ceramic Society*, vol. 37, nº 13, p. 3927-3933, oct. 2017, doi: 10.1016/j.jeurceramsoc.2017.04.058.

[53] H. Zaitouni, L. Hajji, D. Mezzane, E. Choukri, A.G. Razumnaya, Y. Gagou, K. Hoummada, A. Alimoussa, B. Rožič, D. Črešnar, M. El Marssi, et Z. Kutnjak, « Enhanced electrocaloric and energy-storage properties of environment-friendly ferroelectric $Ba_{0.9}Sr_{0.1}Ti_{1-x}Sn_xO3$ ceramics », *Materials Today Communications*, vol. 31, p. 103351, juin 2022, doi: 10.1016/j.mtcomm.2022.103351.

[54] T. Badapanda, S. Chaterjee, A. Mishra, R. Ranjan, et S. Anwar, « Electric field induced strain, switching and energy storage behaviour of lead free Barium Zirconium Titanate ceramic », *Physica B: Condensed Matter*, vol. 521, p. 264-269, sept. 2017, doi: 10.1016/j.physb.2017.07.013.

[55] Q. Xu, M.T. Lanagan, X. Huang, J. Xie, L. Zhang, H. Hao, et H. Liu, « Dielectric behavior and impedance spectroscopy in lead-free BNT–BT–NBN perovskite ceramics for energy storage », *Ceramics International*, vol. 42, nº 8, p. 9728-9736, juin 2016, doi: 10.1016/j.ceramint.2016.03.062.

**Table captions**

Table 1: Structural parameters of BTSn, BCZT, and 0.4BCZT-0.6BTSn obtained from Rietveld refinement.

Table 2: Comparison of energy storage performances of 0.4BCZT–0.6BTSn ceramic with other lead-free ferroelectric ceramics.

**Figure captions**

Figure 1: Room temperature XRD patterns of BTSn, BCZT and 0.4BCZT–0.6BTSn ceramics.

Figure 2: Rietveld refinement of (a) BTSn, (b) BCZT, and (c) 0.4BCZT-0.6BTSn samples. (d) Williamson-Hall plots of BTSn, BCZT, and 0.4BCZT-0.6BTSn samples.

Figure 3: SEM images of (a) BCZT, (b) BTSn and (c) 0.4BCZT–0.6BTSn ceramics.

Figure 4: Temperature dependence of dielectric constant and dielectric loss at various frequencies : (a-b) BCZT, (c-d) BTSn and (e-f) 0.4BCZT–0.6BTSn.

Figure 5: Temperature dependence of the $1/\varepsilon$ at 1 kHz (a, b and c), the relationships between $\ln(1/\varepsilon - 1/\varepsilon_m)$ and $\ln(T - T_m)$ (d, e and f) for all investigated ceramics.

Figure 6: P-E hysteresis loops at different temperatures of (a) BCZT, (b) BTSn and (c) 0.4BCZT-0.6BTSn. (d) P-E hysteresis loops at RT of different samples. (e) $P_r$ and $E_c$ values of all samples.

Figure 7: Schematic diagram of the energy storage calculation of the energy storage parameters.

Figure 8: Temperature dependence of (a) recoverable energy density and (b) energy storage efficiency. (c) Thermal stability of energy storage efficiency.

**Table 1**

| Sample | Unit Cell Parameters (Å) | | | $\chi^2$ | Strain (10$^{-3}$) | Synthesis method |
|---|---|---|---|---|---|---|
| | Phase 1 : *P4mm* | Phase 2 : *Pmm*2 | Phase 3 : *Amm*2 | | | |
| BTSn | a = b = 4.0155<br>c = 4.0197<br>α = β = γ = 90°<br>c/a=1.00104 | - | - | 1.46 | 1.2 | Sol-gel |
| BCZT | a = b = 4.0146<br>c = 4.0111<br>α = β = γ = 90°<br>c/a=0.9991 | a = 3.9969<br>b = 4.0143<br>c = 4.0076<br>α = β = γ = 90°<br>c/a=1.00267 | - | 2.78 | 2.1 | Sol-gel |
| 0.4BCZT-0.6BTSn | a = b = 4.0129<br>c = 4.0069<br>α = β = γ = 90°<br>c/a=0.9985 | - | a = 4.0105<br>b = 4.0098<br>c = 4.0113<br>α = β = γ = 90<br>c/a= 1.00019 | 4.79 | 1.8 | Sol-gel |

**Table 2**

| Samples | $W_{rec}$ (mJ/cm$^3$) | E (kV/cm) | $\eta$ (%) | T (°C) | Refs. |
|---|---|---|---|---|---|
| **BCZT** | **99.88** | **30** | **74.12** | **100.71** | **This work** |
| **BTSn** | **99.83** | **30** | **55.73** | **44.75** | **This work** |
| **0.4BCZT–0.6BTSn** | **137.86** | **30** | **86.19** | **80** | **This work** |
| Ba$_{0.85}$Ca$_{0.15}$Zr$_{0.10}$Ti$_{0.90}$O$_3$ | 62 | 25 | 72.9 | 130 | [21] |
| Ba$_{0.85}$Ca$_{0.15}$Zr$_{0.10}$Ti$_{0.90}$O3 | 14 | 6.5 | 80 | 129 | [49] |
| BaTi$_{0.89}$Sn$_{0.11}$O$_3$ | 72.4 | 25 | 85.07 | 30 | [17] |
| 0.96Bi$_{0.50}$(Na$_{0.80}$K$_{0.20}$)$_{0.50}$TiO$_3$–0.04Ba$_{0.90}$Ca$_{0.10}$Ti$_{0.90}$Zr$_{0.10}$O$_3$ | 373 | 45 | - | 50 | [51] |
| BaCe$_{0.15}$Ti$_{0.85}$O$_3$ | 115 | 24 | 65 | - | [52] |
| Ba$_{0.90}$Sr$_{0.10}$TiO$_3$ | 24 | 7.4 | 85 | 115 | [53] |
| Na$_{0.5}$(Bi$_{0.98}$Dy$_{0.02}$)$_{0.5}$TiO$_3$ | 1200 | 90 | 65 | 200 | [50] |
| BaZr$_{0.05}$Ti$_{0.95}$O$_3$ | 218 | 50 | 72 | RT | [54] |
| 0.85[(1- x)Bi$_{0.5}$Na$_{0.5}$TiO$_3$–xBaTiO$_3$]–0.15Na$_{0.73}$Bi$_{0.09}$NbO$_3$ | 1100 | 122 | 67.9 | - | [55] |

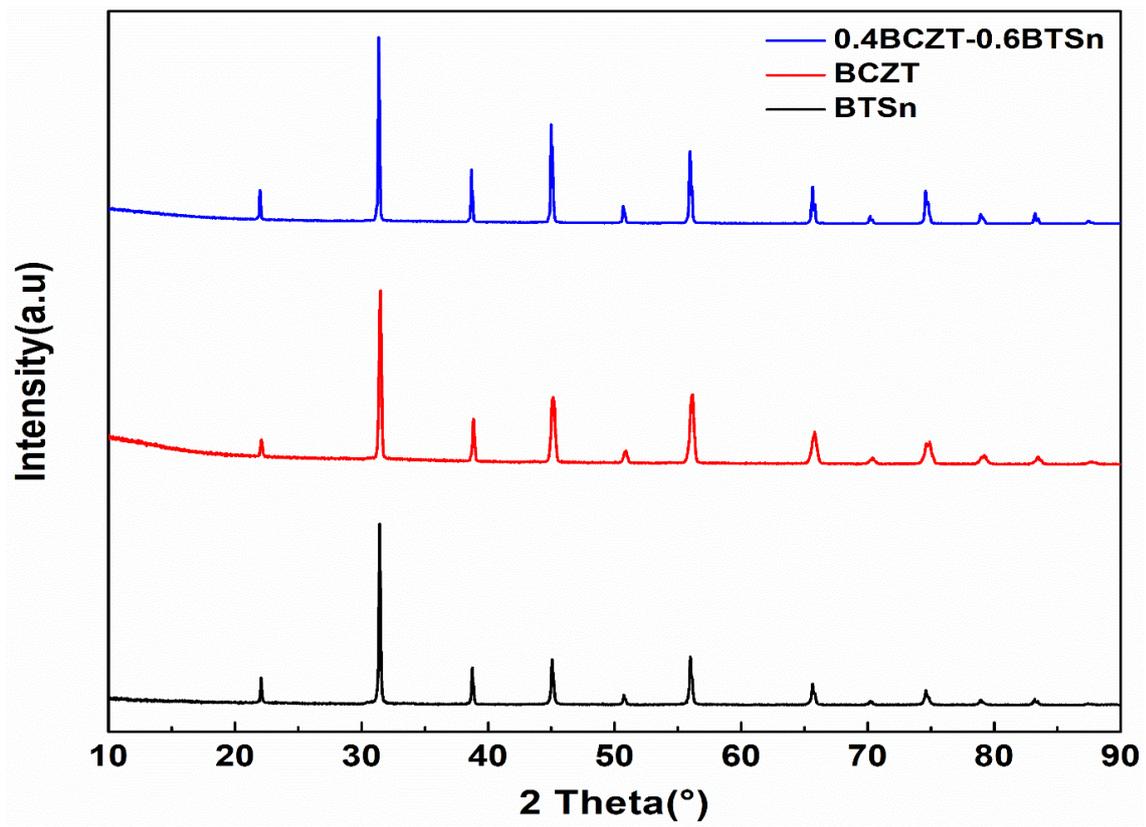

Figure 1

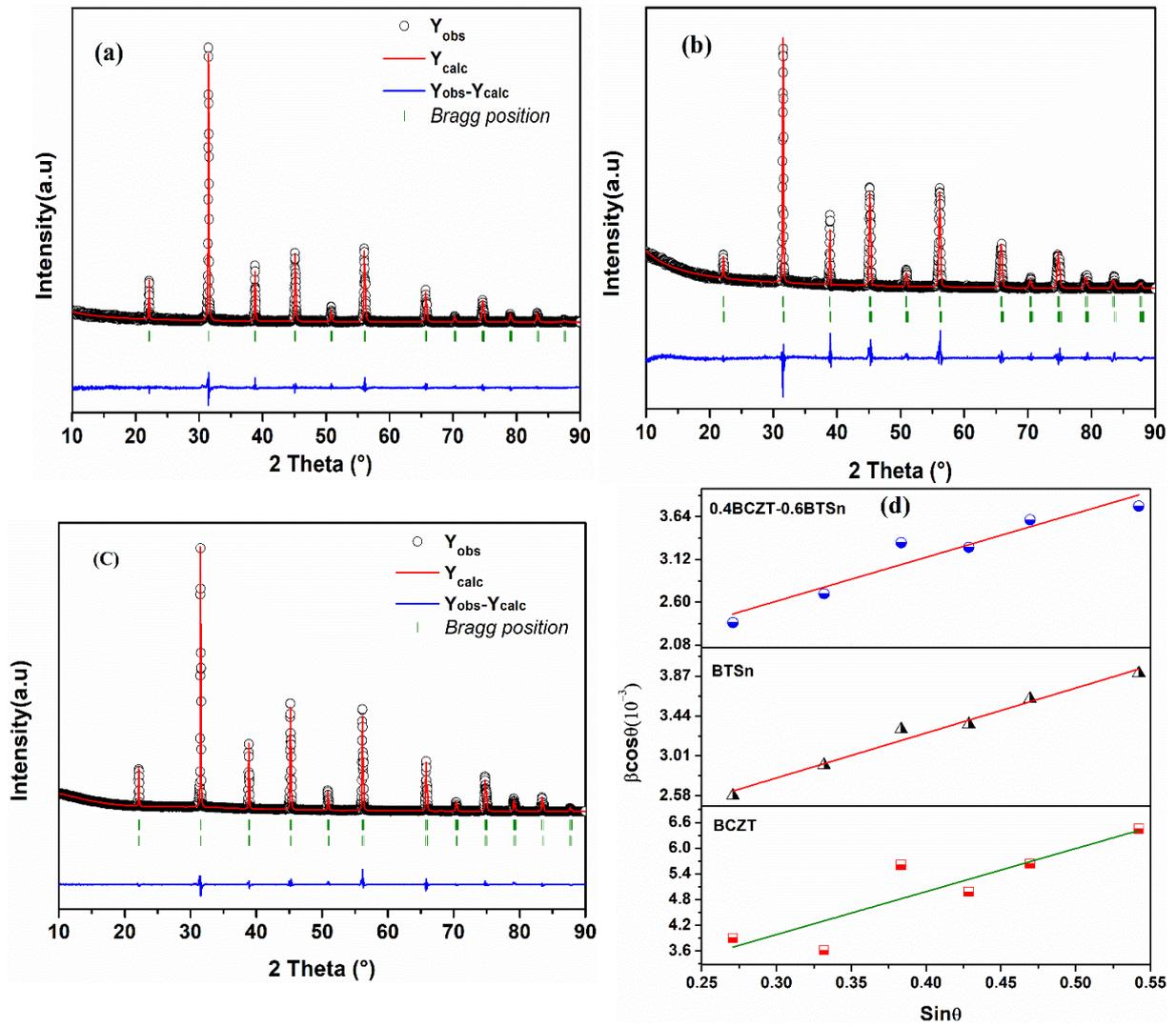

Figure 2

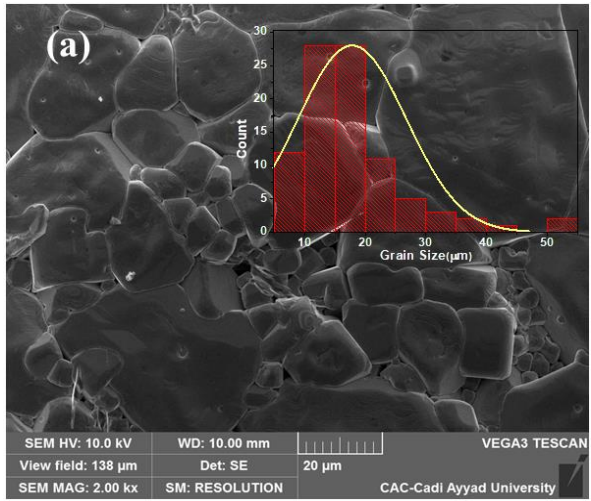
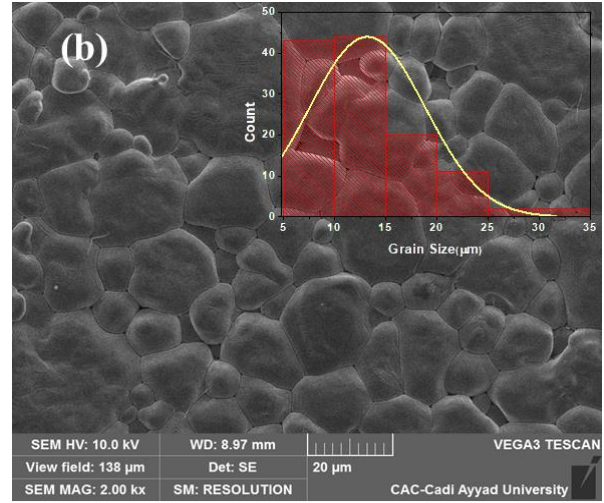
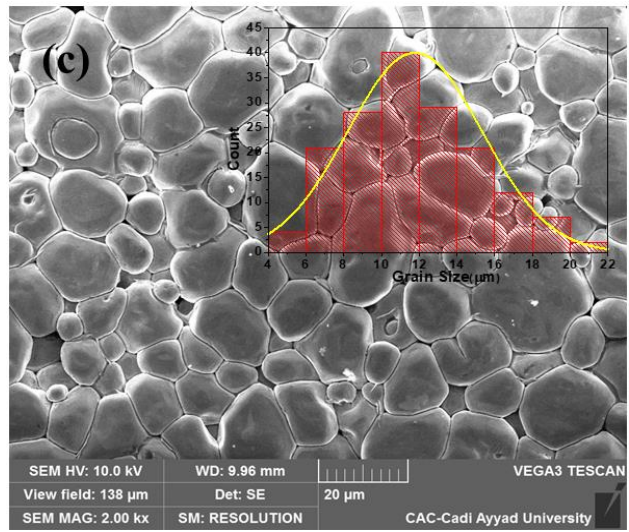

Figure 3

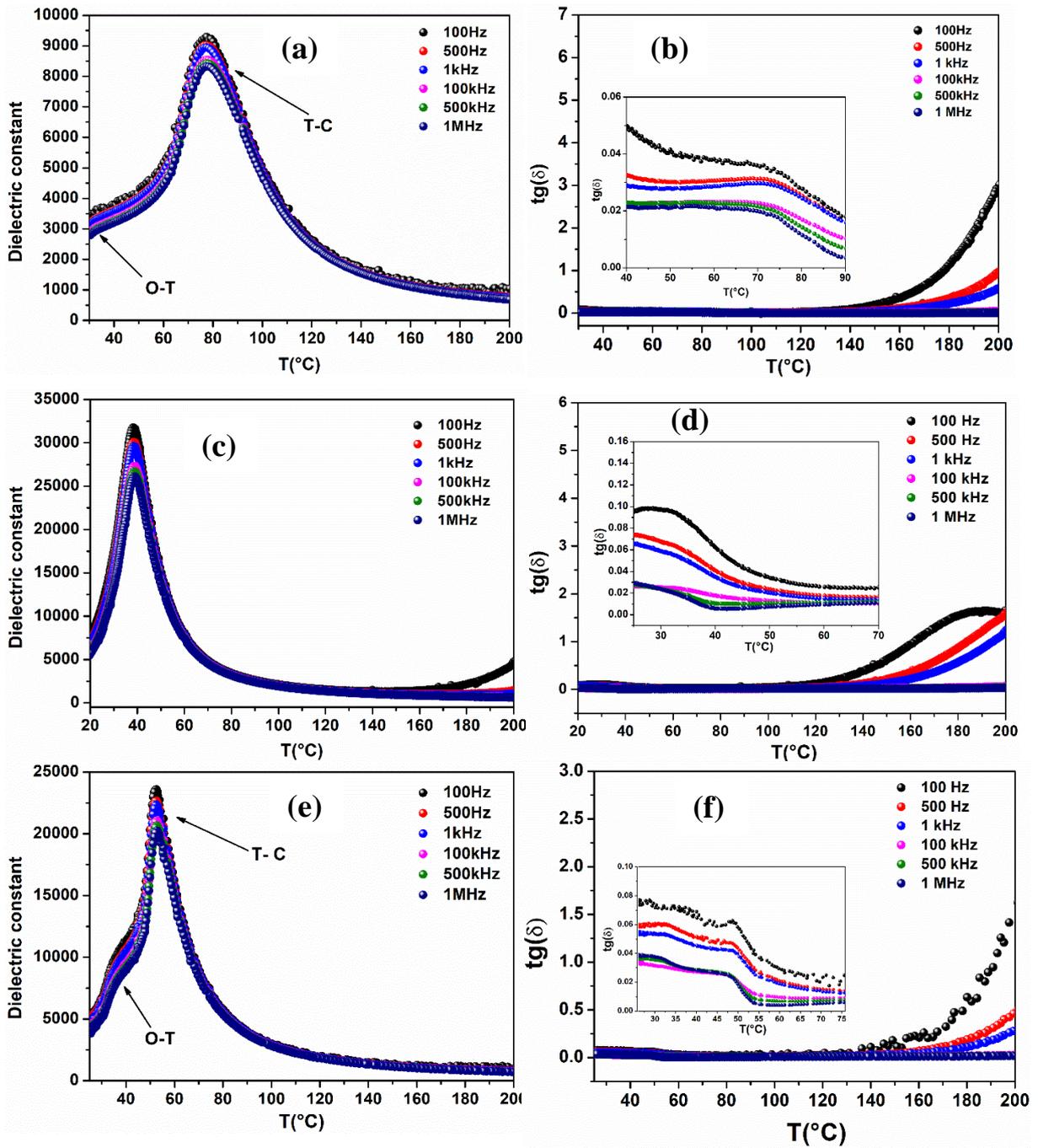

Figure 4

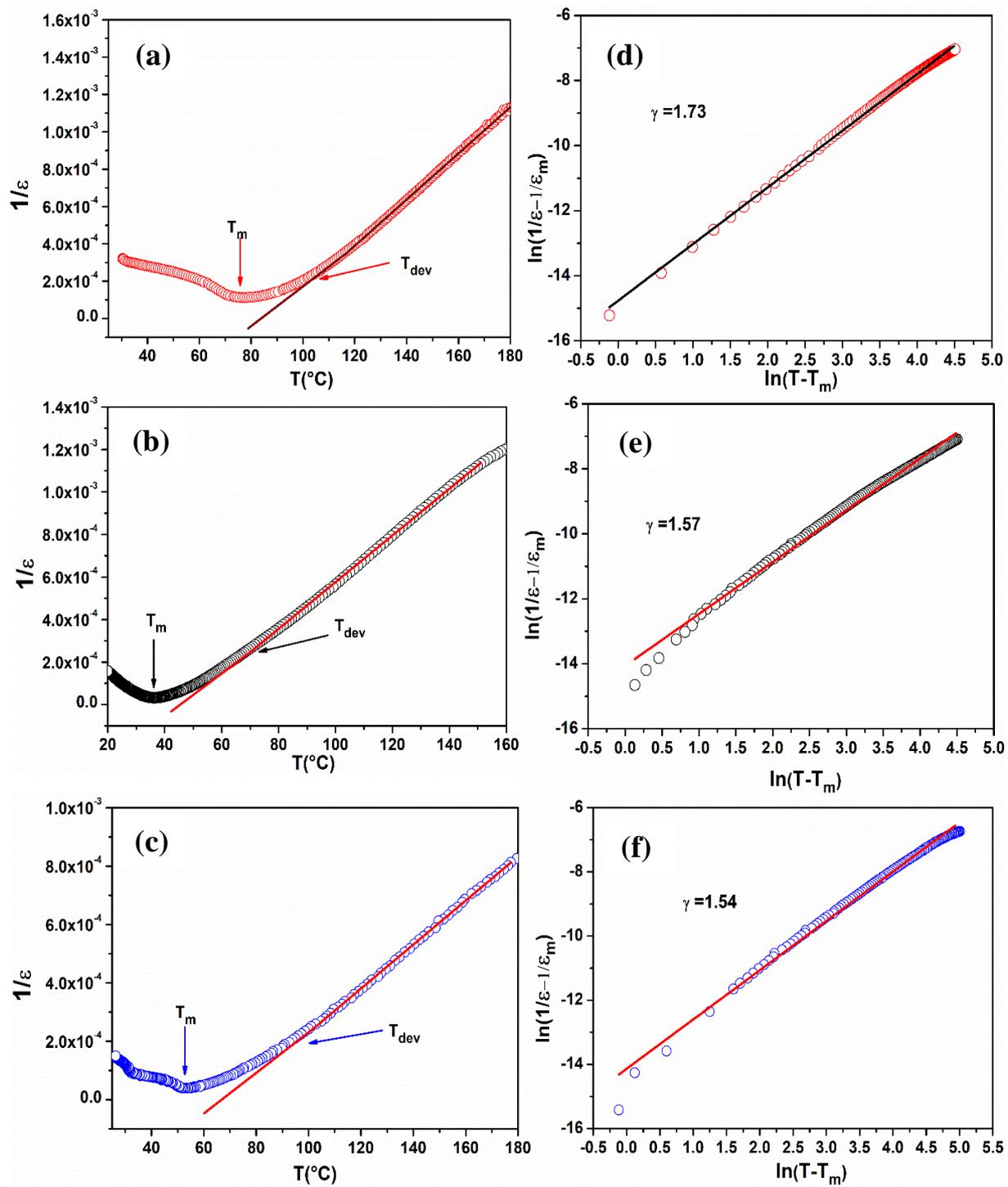

Figure 5

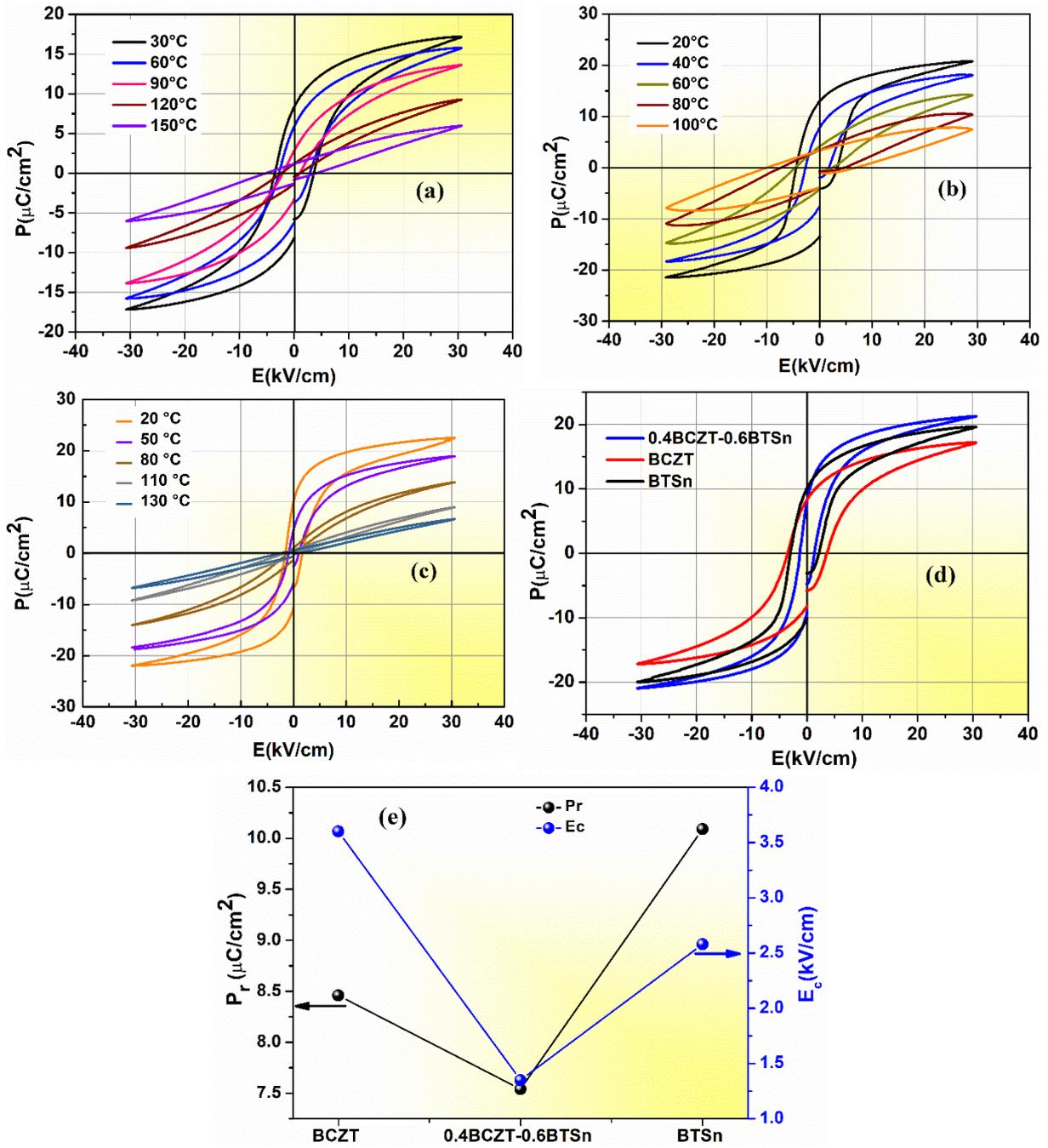

Figure 6

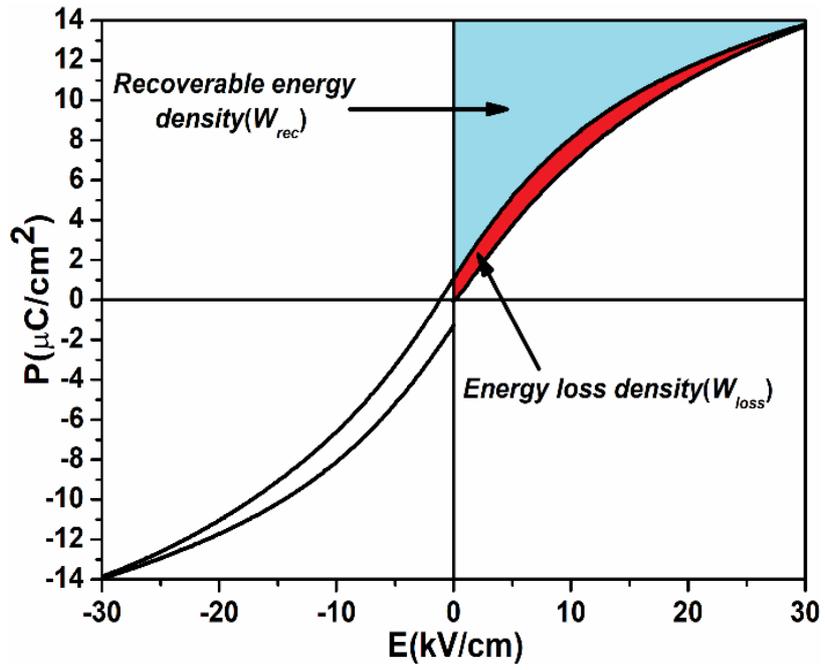

Figure 7

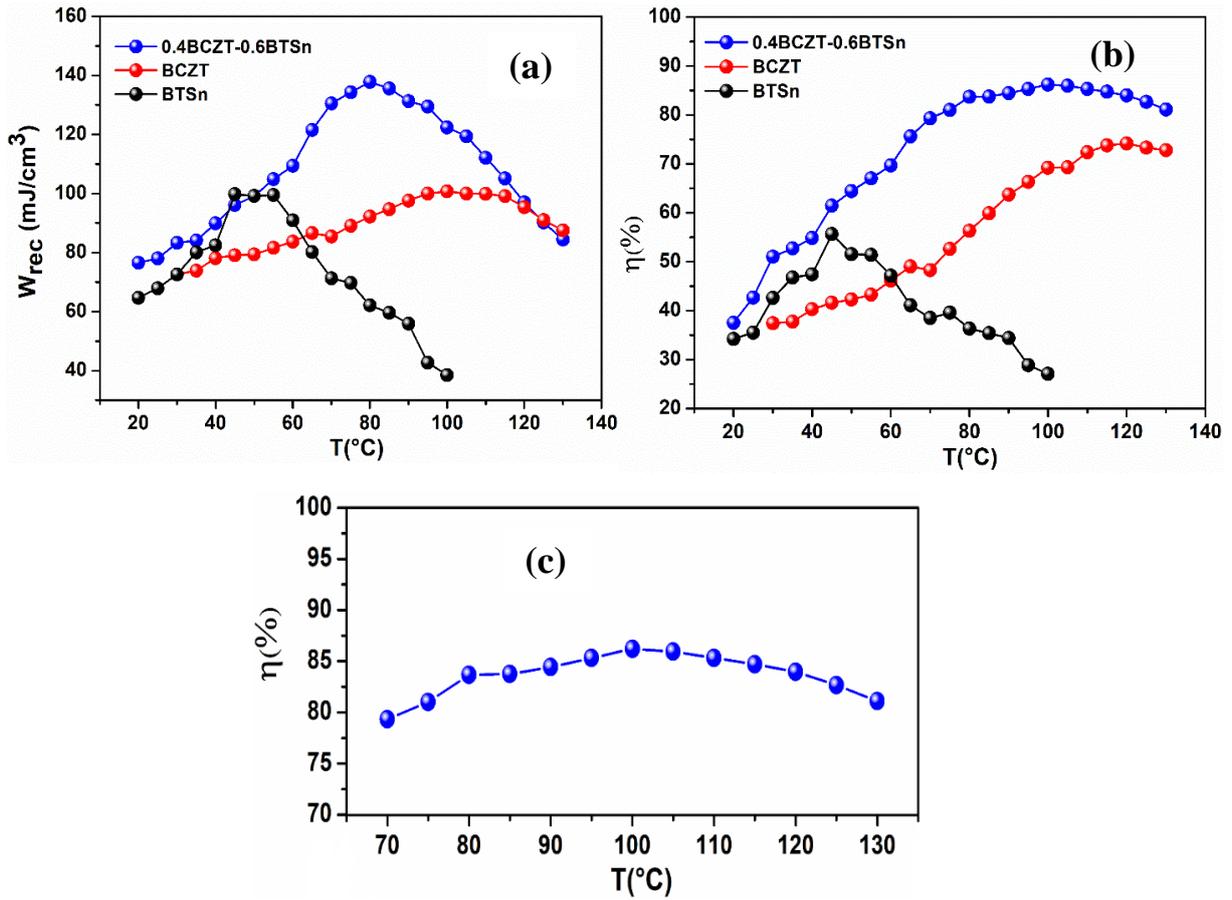

Figure 8